\documentclass[11pt]{article}
\usepackage{coling2016}
\usepackage{amsmath}
\usepackage{xcolor}
\usepackage{multirow}
\usepackage{times}
\usepackage{helvet}
\usepackage{url}
\usepackage{graphicx}
\usepackage{courier}
\graphicspath{{figures/}}
\usepackage{comment}
\usepackage{enumitem}
\usepackage{fancyvrb}
\usepackage{color}
\usepackage{subcaption}

\begin{document}

\title{OCR++: A Robust Framework For Information Extraction from Scholarly Articles}

\author{Mayank Singh, Barnopriyo Barua, Priyank Palod, Manvi Garg,\\
 \bf{Sidhartha Satapathy, Samuel Bushi, Kumar Ayush, Krishna Sai Rohith, Tulasi Gamidi,}\\
\bf{Pawan Goyal and Animesh Mukherjee}\\
Department of Computer Science and Engineering\\
Indian Institute of Technology, Kharagpur, WB, India\\
mayank.singh@cse.iitkgp.ernet.in\\
}
\date{15 October 2015}

\maketitle

\begin{abstract}
This paper proposes \textit{OCR++}, an open-source framework designed for a variety of 
information extraction tasks from scholarly articles including metadata (title, author names, affiliation and e-mail), 
structure (section headings and body text, table and figure headings, URLs and footnotes) and bibliography (citation instances and references).
We analyze a diverse set of scientific articles written in English language to understand generic writing patterns and formulate rules to develop this hybrid framework.
Extensive evaluations show that the proposed framework outperforms the existing state-of-the-art tools with huge margin in structural information
extraction along with improved performance in metadata and bibliography extraction tasks, both in terms of accuracy (around 50\% improvement) and processing time (around 52\% improvement). 
A user experience study conducted with the help of 30 researchers reveals that the researchers found this system to be very helpful. As an additional objective,
we discuss two novel use cases including automatically extracting links to public datasets from the proceedings, which would further accelerate the
advancement in digital libraries. The result of the framework can be exported as a whole into structured TEI-encoded documents. 
Our framework is accessible online at \textit{http://www.cnergres.iitkgp.ac.in/OCR++/home/}.
\end{abstract}

\section{Introduction}
Obtaining structured data from documents is necessary to support retrieval 
tasks~\cite{beel2011introducing}. Various scholarly organizations and companies deploy information extraction tools in 
their production environments. Google scholar\footnote{http://scholar.google.com}, Microsoft academic 
search\footnote{http://academic.research.microsoft.com}, Researchgate\footnote{researchgate.net}, 
CiteULike\footnote{http://www.citeulike.org/} etc. provide academic search engine facilities. 
European publication server (EPO)\footnote{https://data.epo.org/publication-server/?lg=en}, ResearchGate and 
Mendeley\footnote{https://www.mendeley.com/} use \textit{GROBID}~\cite{lopez2009grobid} for header extraction and
analysis.  A similar utility named~\textit{SVMHeaderParse} is deployed by $CiteSeer^x$\footnote{http://citeseerx.ist.psu.edu/} for header extraction. 

Through a comprehensive literature survey, we find comparatively less research in document structure analysis than metadata and bibliography extraction from scientific documents. 
The main challenges lie in the inherent errors in OCR processing and diverse formatting styles adopted by different publishing venues. 
 We believe that a key strategy to tackle this problem is to analyze research articles from different publishers to identify generic 
 patterns and rules, specific to various information extraction tasks. We introduce \textit{OCR++}, a hybrid framework to extract textual 
 information such as (i) metadata  -- title, author names, affiliation and e-mail, (ii) structure -- section headings and body text, table and figure headings, URLs and footnotes and (iii) bibliography -- citation instances and references from scholarly articles. The framework employs a variety of Conditional Random Field (CRF) models and hand-written rules specially crafted for handling various tasks. Our framework produce comparative results in metadata extraction tasks. However, it significantly outperform state-of-the-art systems in structural information extraction tasks. On an average, we record an accuracy improvement of 50\% and a processing time improvement of 52\%. We claim that our hybrid approach leads to higher performance than complex machine learning models based systems. 
 We also present two novel use cases including extraction of public dataset links available in the proceedings of the NLP conferences available from the ACL anthology.

\section{Framework overview}
\label{tasks}
\begin{figure}[!thb]
\vspace{-0.2cm}
  \centering
  \resizebox{1\linewidth}{!}{\includegraphics{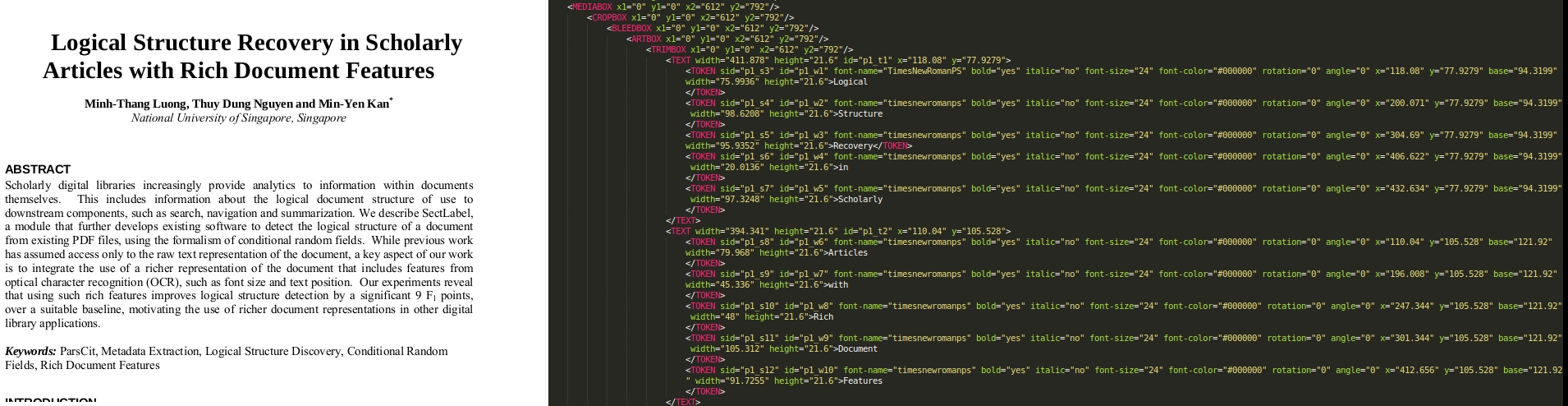}}
  \caption{Screenshot of \textit{pdf2xml} tool output.} \label{fig:pdftoxml}
  \vspace{-0.2cm}
\end{figure}

\begin{figure}[!thb]
\vspace{-0.2cm}
\centering
\begin{subfigure}{.48\textwidth}
  \centering
  \includegraphics[width=1\linewidth]{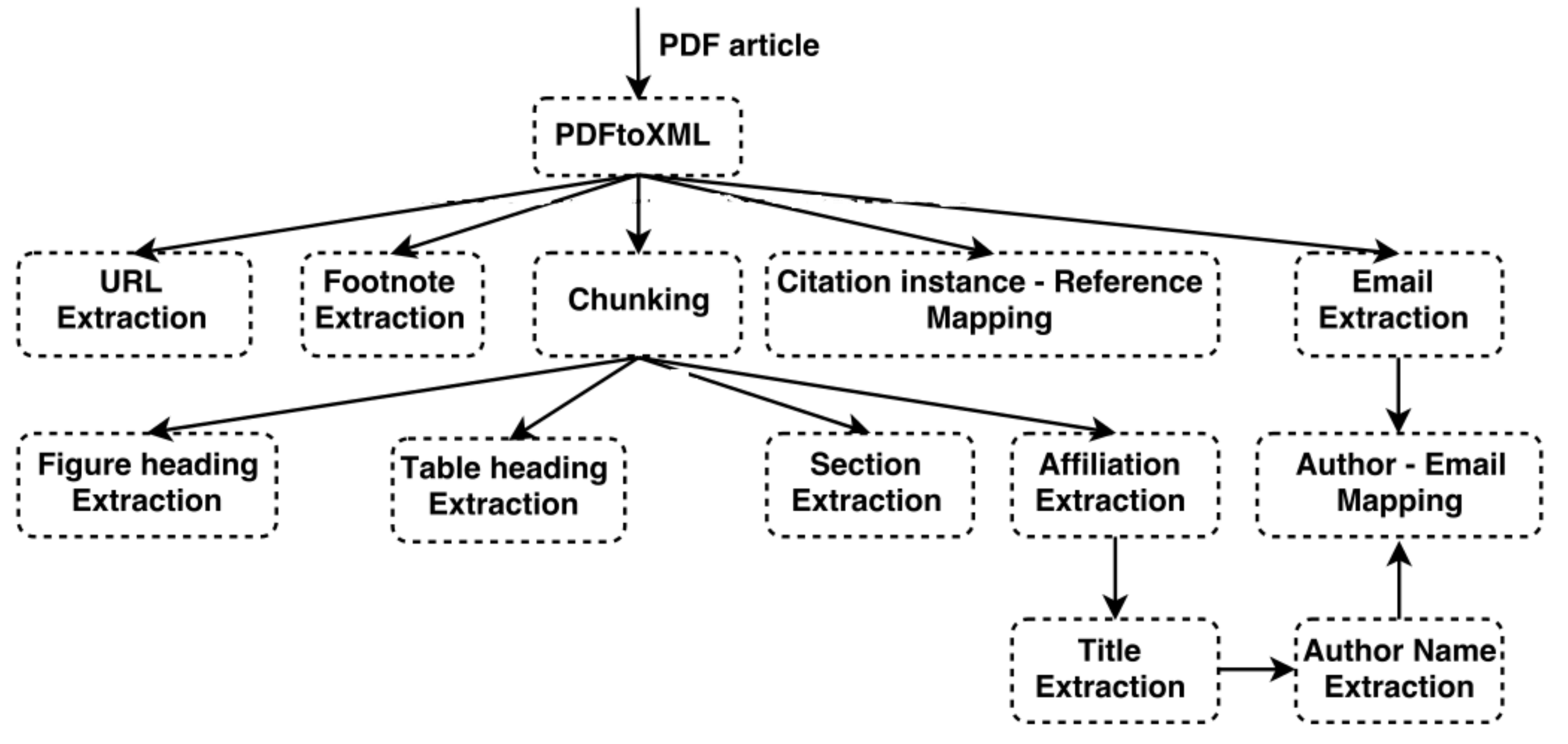}
  \caption{Sub-tasks dependencies in \textit{OCR++}.}
  \label{fig:sub1}
\end{subfigure}
\begin{subfigure}{.5\textwidth}
  \centering
  \includegraphics[width=1\linewidth]{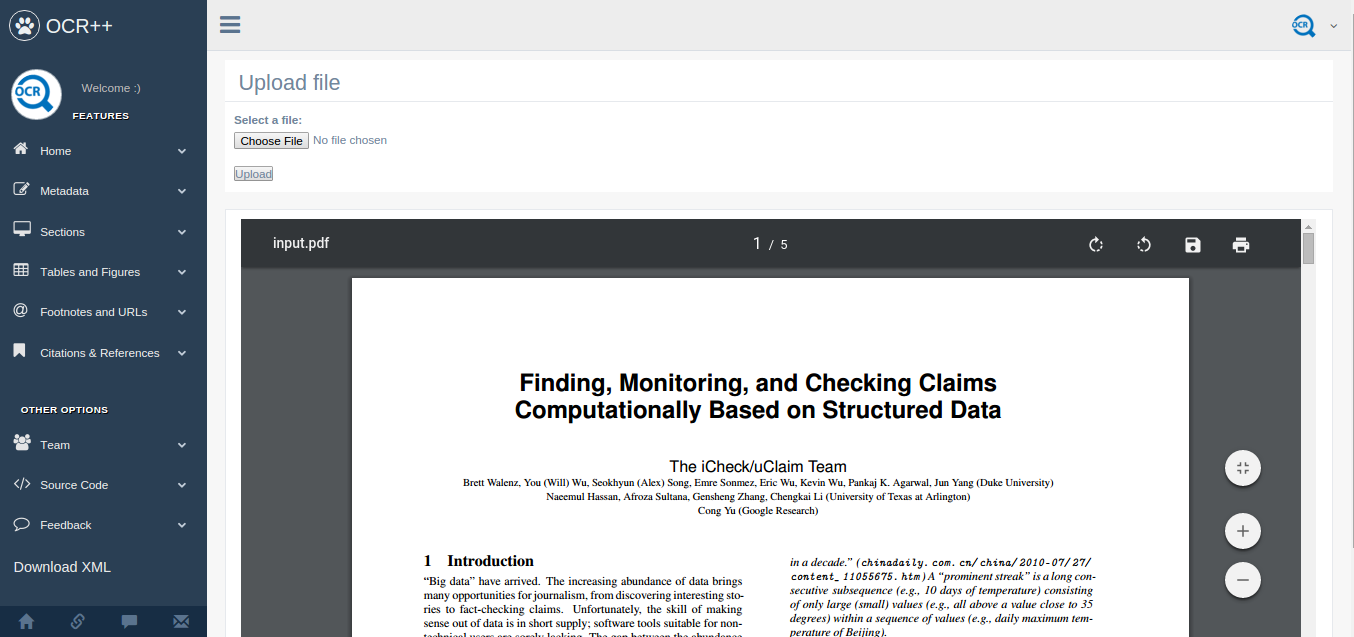}
  \caption{Screenshot of OCR++ web interface.}
  \label{fig:sub2}
\end{subfigure}
\caption{OCR++ framework overview and user interface}
\label{fig:ocr_interface}
\vspace{-0.2cm}
\end{figure}

\textit{OCR++} is an extraction framework for scholarly articles, completely written in Python (Figure~\ref{fig:ocr_interface}).
The framework takes a PDF article as input, 1) converts the PDF file to an XML format, 2) processes the XML file to 
extract useful information, and 3) exports output in structured TEI-encoded\footnote{http://www.tei-c.org/index.xml} documents. 
We use \textit{pdf2xml}\footnote{http://sourceforge.net/projects/pdf2xml/} to convert PDF files into rich XML files.
Each token in the PDF file is annotated with rich metadata, namely, $x$ and $y$ co-ordinates, font size, font weight, font style 
etc. (Figure \ref{fig:pdftoxml}).

Figure~\ref{fig:ocr_interface}(a) describes the sub-task dependencies in \textit{OCR++}. The web interface of the tool is shown in Figure~\ref{fig:ocr_interface}(b).
We leverage the rich information present in the XML files to perform extraction tasks. 
Although each extraction task described below is performed using machine learning models as well as hand written rules/heuristics, 
we only include the better performing scheme in our framework. Next, we describe each extraction task in detail.

\vspace{-0.1cm}
\subsection{Chunking}
\label{sec:chunking}
\vspace{-0.1cm}
As a first step, we segment XML text into \textit{chunks} by measuring distance from neighboring text and differentiating from the 
surrounding text properties such as font-size and bold-text.

\vspace{-0.1cm}
\subsection{Title extraction}
\label{sec:title}
\vspace{-0.1cm}
We train a CRF model to label the token sequences using 6300 training instances. Features are constructed based on generic
characteristics of formatting styles. Token level feature set includes boldness, relative position in the paper, relative position in the first chunk, 
relative size, case of first character, boldness + relative font size overall, case of first character in present and next token and case of first character in present and previous token.

\vspace{-0.1cm}
\subsection{Author name extraction}
\label{sec:author_name}
\vspace{-0.1cm}
In this sub-task, we use the same set of features as described in the title extraction sub-task to learn the CRF model along with a 
heuristic that the tokens eligible for the author names are either present in the first section or within 120 tokens after the title. 
Different author names are distinguished using heuristics, such as, difference in $y$-coordinates, tab separation etc.
Further, false positives are removed using heuristics such as, length of consecutive author name tokens, symbol or digit in token and POS tag.
The first word among consecutive tokens is considered as the first name, the last word as the last name, and all the 
remaining words are treated together as the middle name.

\vspace{-0.1cm}
\subsection{Author e-mail extraction}
\vspace{-0.1cm}
An e-mail consists of a user name, a sub-domain name and a domain name. In case of scholarly articles, usually, the user names are written inside brackets 
separated by commas and the bracket is succeeded by the sub-domain and domain name. On manual 
analysis of a set of scholarly articles, we find four different writing patterns, author1@cse.domain.com, \{author1, author2, author3\}@cse.domain.com, 
[author4, author5, author6]@cse.domain.com and [author7@cse, author7@ee].domain.com. Based on these observations, we construct hand written rules to extract e-mails.

\vspace{-0.1cm}
\subsection{Author affiliation extraction}
\vspace{-0.1cm}
We use hand written rules to extract affiliations. We employ heuristics such as, presence of country name, tokens like ``University", ``Research", ``Laboratories", ``Corporation", ``College", ``Institute", superscript character etc.

\vspace{-0.1cm}
\subsection{Headings and section mapping}
\label{sec:sec_mapping}
\vspace{-0.1cm}
We employ CRF model to label section headings. Differentiating features (the first token of the chunk, the second token, avg. boldness of the chunk, avg. font-size, Arabic/Roman/alpha-enumeration etc.) are extracted from chunks to train the CRF.

\vspace{-0.1cm}
\subsection{URL}
\vspace{-0.1cm}
We extract URLs using a single regular expression described below:
\vspace{-0.1cm}
\begin{Verbatim}[fontsize=\small]
http[s]?://(?:[a-zA-Z]|[0-9]|[\$-_@.&+]|[!*\(\),]|(?:\%[0-9a-fA-F][0-9a-fA-F]))
\end{Verbatim}
\vspace{-0.1cm}
\vspace{-0.1cm}
\subsection{Footnote}
\vspace{-0.1cm}
Most of the footnotes have numbers or special symbols (like asterisk etc.) at the beginning in the 
form of a superscript. Footnotes have font-size smaller than the surrounding text and are found in the lower end of a page --
average font size of tokens in a chunk and $y$-coordinate were used as features for training the CRF. Moreover, footnotes are found in 
the lower half of the page (this heuristic helped in filtering false positives).

\vspace{-0.1cm}
\subsection{Figure and table headings}
\vspace{-0.1cm}
Figure and table heading extraction is performed after chunking (described in Section \ref{sec:chunking}).
If the chunk starts with the word ``FIGURE" or ``Figure" or ``FIG." or ``Fig.", then the chunk represents a figure heading. Similarly, 
if the chunk starts with the word ``Table" or ``TABLE", then the chunk represents a table heading. However, it has been observed that table
contents are also present in the chunk. Therefore, we use a feature ``bold font" to extract bold tokens from such chunks. 

\vspace{-0.1cm}
\subsection{Citations and references}
\vspace{-0.1cm}
The bibliography extraction task includes extraction of citation instances and references. All 
the tokens succeeding the reference section are considered to be part of references and further each reference is extracted separately. Again, we employ
hand written rules to distinguish between two consecutive references. On manual analysis, we found 16 unique citation instance writing styles (Table \ref{tab:citation_regular_expresssion}). 
We code these styles into regular expressions to extract citation instances.

\begin{table*}[!thb]
\vspace{-0.3cm}
\small
 \centering
 \caption{Generic set of regular expressions for citation instance identification. Here, AN represent author name, Y represent year and I represent reference index within citation instance.}
 \begin{tabular}{|l|l|}
  \hline
 \textbf{Citation Format}& \textbf{Regular Expression}\\\hline
$<$AN$>$ et al. $[<$I$>]$ & \verb/([A-Z][a-zA-Z]* et al[.][\string\d\{1,3\}])/\\\hline
$<$AN$>$ $[<$I$>]$&\verb/([A-Z][a-zA-Z]* [\string\d\{2\}])/\\\hline
$<$AN$>$ et al.$<$spaces$>[<$I$>]$&\verb/([A-Z][a-zA-Z]* et al[.][ ]*[\string\d\{1\}])/\\\hline
$<$AN$>$ et al., $<$Y$><$I$>$&\verb/([A-Z][a-zA-Z]* et al[.],\string\d\{4\}[a-z])/\\\hline
$<$AN$>$ et al., $<$Y$>$&\verb/([A-Z][a-zA-Z]* et al[.][,] \string\d\{4\})/\\\hline
$<$AN$>$ et al., $(<$Y$>)$&\verb/([A-Z][a-zA-Z]* et al[.][,] (\string\d\{4\}))/\\\hline
$<$AN$>$ et al. $<$Y$>$&\verb/([A-Z][a-zA-Z]* et al[.] \string\d\{4\})/\\\hline
$<$AN$>$ et al. $(<$Y$>)$&\verb/([A-Z][a-zA-Z]* et al[.] (\string\d\{4\}))/\\\hline
$<$AN$>$ and $<$AN$>$ $(<$Y$>)$&\verb/([A-Z][a-zA-Z]* and [A-Z][a-zA-Z]*(\string\d\&{4\}))/\\\hline
$<$AN$>$ \& $<$AN$>$ $(<$Y$>)$&\verb/([A-Z][a-zA-Z]* & [A-Z][a-zA-Z]*(\string\d\&{4\}))/\\\hline
$<$AN$>$ and $<$AN$>$, $<$Y$>$&\verb/([A-Z][a-zA-Z]* and [A-Z][a-zA-Z]*[,] \d\{4\})/\\\hline
$<$AN$>$ \& $<$AN$>$, $<$Y$>$&\verb/([A-Z][a-zA-Z]* & [A-Z][a-zA-Z]*[,] \d\{4\})/\\\hline
$<$AN$>$, $<$Y$>$&\verb/([A-Z][a-zA-Z]*[,] \string\d\{4\})/\\\hline
$<$AN$>$ $<$Y$>$&\verb/([A-Z][a-zA-Z]* \string\d\{4\})/\\\hline
$<$AN$>$, $(<$Y$><$I$>)$&\verb/([A-Z][a-zA-Z]*(\string\d\{4\}[a-z]*))/\\\hline
$<$ multiple indices seperated by commas $>$&\verb/.*?[(.*?)]/\\\hline
\end{tabular}
\label{tab:citation_regular_expresssion}
\vspace{-0.4cm}
\end{table*}

\subsection{Mapping tasks}

\noindent\textbf{Connecting author name to e-mail:} In general, each author name present in the author section associates with some e-mail. 
\textit{OCR++} tries to recover this association using simple rules, for example, sub-string match between username and author names,
abbreviated full name as username, order of occurrence of e-mails etc.

\noindent\textbf{Citation reference mapping:} Each extracted citation instance is mapped to respective reference. Since, there are two different styles of writing  citation instances, 
\textit{Indexed} and \textit{Non-indexed}, we define mapping tasks for each style separately. Indexed citations are mapped directly to references
with the index inside enclosed brackets. The extracted index is mapped with the corresponding reference. Non-indexed citations are represented
using combination of year of publication and author's last name.

\vspace{-0.1cm}
\section{Results and discussion}
\label{sec:evaluation}
\vspace{-0.1cm}
Following an evaluation carried out by \newcite{lipinski2013evaluation}, GROBID provided the best results over seven existing systems, with several metadata
recognized with over 90\% precision and recall. Therefore, we compare \textit{OCR++} with the state-of-the-art \textit{GROBID}.
We compare results for each of the sub-tasks for both the systems against the ground-truth dataset. The ground-truth dataset is prepared by manual
annotation of title, author names, affiliations, URLs, sections, subsections, section headings, table headings, figure headings and references for 
138 articles from different publishers. The publisher names are present in Table~\ref{tab:publisher-level}. We divide article set into training and test datasets in the ratio of 20:80. 
Note that each of the extraction modules described in the previous section also have separate training sample count, for instance, 6300 samples have been used to train the title extraction. Also, we observe 
that both the systems provide partial results in some cases. For example, in some cases, only half of the title is extracted 
or the author names are incomplete. In order to accommodate partial results from extraction tasks, we provide evaluation results at token 
level, i.e, what fraction of the tokens are correctly retrieved.

Table~\ref{tab:subtask-level} presents comparative results for GROBID and \textit{OCR++}. It shows that in terms of precision, \textit{OCR++} 
outperforms \textit{GROBID} in all the sub-tasks. Recall is higher for \textit{GROBID} for some of the metadata extraction tasks. 
In \textit{OCR++}, since title extraction depends on the first extracted chunk from section extraction, the errors in chunk extraction lead to 
low recall in title extraction. Similar problem results in lower recall in author name extraction.  Due to the presence of variety of white space length between author first, middle and last name in various formats, we observe low recall overall in author name extraction subtasks. We also found that in many cases author-emails are quite different from author names resulting in lower recall for author-email extraction subtask. \textit{OCR++} outperforms \textit{GROBID} 
in majority of the structural information extraction subtasks in terms of both precision and recall. We observe that \textit{GROBID} performs poorly 
for table heading extraction due to intermingling of table text with heading tokens and unnumbered footnotes.
Similar argument holds for the figure heading as well. URL extraction feature is not implemented in \textit{GROBID}, while \textit{OCR++} extracts
it very accurately. Similarly, poor extraction of non-indexed footnotes resulted in lower recall for footnote extraction subtask.   

Similarly, Table \ref{tab:publisher-level} compares \textit{GROBID} and \textit{OCR++} 
for different publishing formats. Here the results seem to be quite impressive with \textit{OCR++} outperforming \textit{GROBID} in almost
all cases. This demonstrates the effectiveness and robustness of using generic patterns and rules used in building \textit{OCR++}. As our system is more biased towards single and double column formats, we observe less performance on three column formats. Similarly, non-indexed sections format show less performance than  indexed sections format. 

\begin{table}[!thb]
\vspace{-0.4cm}
\centering
  \caption{Micro-average accuracy for GROBID and OCR++ for different extractive subtasks.}
   \resizebox{0.8\textwidth}{!}{\begin{minipage}{\textwidth}
  \begin{tabular}{l|c|c|c||c|c|c}
  \hline
\multirow{3}{*}{Subtask} & \multicolumn{3}{|c||}{GROBID} & \multicolumn{3}{|c}{OCR++}\\\cline{2-7}
&Precision&Recall&F-Score&Precision&Recall&F-Score\\\hline
Title &0.93&\textbf{0.94}&\textbf{0.93}& \textbf{0.96}&0.85&0.90\\
Author First Name&0.81&\textbf{0.81}&\textbf{0.81}&\textbf{0.91}&0.65&0.76\\
Author Middle Name&N/A&N/A&N/A&\textbf{1.0}&\textbf{0.38}&\textbf{0.55}\\
Author Last Name&0.83&\textbf{0.82}&\textbf{0.83}&\textbf{0.91}&0.65&0.76\\
Email&0.80&0.20&0.33&\textbf{0.90}&\textbf{0.93}&\textbf{0.91}\\
Affiliation&0.74&0.60&0.66&\textbf{0.80}&\textbf{0.76}&\textbf{0.78}\\
Section Headings&0.70&\textbf{0.87}&\textbf{0.78}&\textbf{0.80}&0.72&0.76\\
Figure headings&0.59&0.42&0.49&\textbf{0.96}&\textbf{0.75}&\textbf{0.84}\\
Table headings&0.77&0.17&0.28&\textbf{0.87}&\textbf{0.74}&\textbf{0.80}\\
URLs&N/A&N/A&N/A&\textbf{1.0}&\textbf{0.94}&\textbf{0.97}\\
Footnotes&0.80&0.42&0.55&\textbf{0.91}&\textbf{0.63}&\textbf{0.91}\\
Author-Email&0.38&0.24&0.29&\textbf{0.93}&\textbf{0.44}&\textbf{0.60}\\\hline
\end{tabular}
 \end{minipage}}
\label{tab:subtask-level}
\vspace{-0.3cm}
\end{table}

\begin{table}[!thb]
\vspace{-0.3cm}
\centering
  \caption{Micro-average accuracy for GROBID and OCR++ for different publishing styles.}
   \resizebox{0.8\textwidth}{!}{\begin{minipage}{\textwidth}
  \begin{tabular}{c|c|c|c|c||c|c|c}
  \hline
\multirow{3}{*}{Publisher} &\multirow{3}{*}{paper count}& \multicolumn{3}{|c||}{GROBID} & \multicolumn{3}{|c}{OCR++}\\\cline{3-8}
&&Precision&Recall&F-Score&Precision&Recall&F-Score\\
\hline
IEEE& 30&0.82&0.61&0.70  & \textbf{0.9}&\textbf{0.69}&\textbf{0.78}\\
ARXIV &25&0.75&0.63&0.68&\textbf{0.91}&\textbf{0.73}&\textbf{0.81}\\
ACM&35&0.69&0.49&0.58&\textbf{0.89}&\textbf{0.71}&\textbf{0.79}\\
ACL&16&0.89&0.59&0.71&\textbf{0.91}&\textbf{0.79}&\textbf{0.85}\\
SPRINGER&17&0.78&0.6&0.68&\textbf{0.85}&\textbf{0.63}&\textbf{0.72}\\
CHI&3&0.13&0.20&0.16&\textbf{0.5}&\textbf{0.36}&\textbf{0.42}\\
ELSEVIER&6&0.58&0.6&0.59&\textbf{0.82}&\textbf{0.74}&\textbf{0.78}\\
NIPS&3&0.82&0.68&0.74&\textbf{0.83}&\textbf{0.72}&\textbf{0.77}\\
ICML&1&\textbf{0.6}&\textbf{0.6}&\textbf{0.6}&0.59&0.54&0.56\\
ICLR&1&0.49&\textbf{0.55}&0.52&\textbf{0.67}&0.52&\textbf{0.59}\\
JMLR&1&0.58&0.55&0.56&\textbf{0.86}&\textbf{0.83}&\textbf{0.85}\\\hline
\end{tabular}
 \end{minipage}}
\label{tab:publisher-level}
\vspace{-0.4cm}
\end{table}

Since citation instance annotation demands significant extent of human-efforts, we randomly select eight PDF articles from eight different publishers from 
ground-truth dataset PDFs. Manual annotation produces 328 citation instances. We 
also annotate references to produce 187 references in total.  
Table~\ref{tab:bibiography_accuracy} shows performance comparison for bibliography related tasks. As depicted from Table 
\ref{tab:bibiography_accuracy}, \textit{OCR++} performs better for both citation and reference extraction tasks. GROBID does
not provide Citation-Reference mapping, which is an additional feature of \textit{OCR++}.

\begin{table}[!thb]
\vspace{-0.4cm}
\centering
  \caption{Micro-average accuracy for GROBID and OCR++ bibliography extraction tasks.}
   \resizebox{0.8\textwidth}{!}{\begin{minipage}{\textwidth}
  \begin{tabular}{c|c|c|c||c|c|c}
  \hline
\multirow{3}{*}{} & \multicolumn{3}{|c||}{GROBID} & \multicolumn{3}{|c}{OCR++}\\\cline{2-7}
&Precision&Recall&F-Score&Precision&Recall&F-Score\\
\hline
Citation&0.93&0.81 &0.87
&\textbf{0.94}&\textbf{0.97}&\textbf{0.95}\\
Reference &0.94&0.94&0.94&\textbf{0.98}&\textbf{0.99}&\textbf{0.98}\\
Citation-Reference&N/A&N/A&N/A&\textbf{0.94}&\textbf{0.97}&\textbf{0.95}\\
\hline
\end{tabular}
 \end{minipage}}
\label{tab:bibiography_accuracy}
\vspace{-0.4cm}
\end{table}

Next, we investigate whether better formatting styles over the years lead to higher precision by the proposed tool.
Also, we compare \textit{OCR++} with \textit{GROBID} in terms of processing time.

\subsection{Effect of formatting style on precision}
\label{sec:time_evolution}
\begin{figure}[!thb]
\vspace{-0.3cm}
  \centering
  \resizebox{.7\linewidth}{!}{\includegraphics{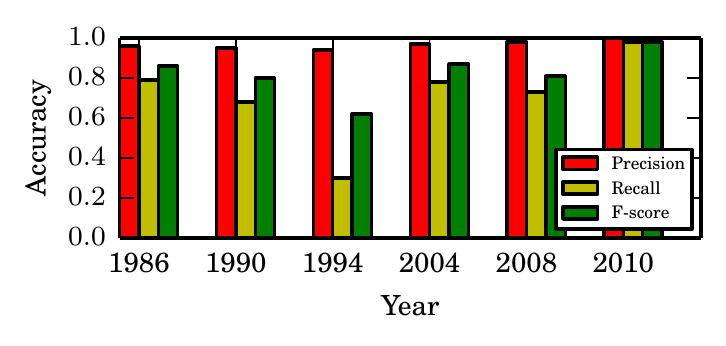}}
  \caption{Title extraction accuracy calculated at six different years for COLING.} \label{fig:time_evolution}
\vspace{-0.5cm}
\end{figure}

We select International Conference on Computational Linguistics (COLING) as a representative example to understand the effect of 
evolution in formatting styles over the years on the accuracy of the extraction task. We select ten random articles each from
six different years of publications. \textit{OCR++} is used to extract title for each year. 
Figure \ref{fig:time_evolution} presents title extraction accuracy for each year, reinstating the fact that the recent year
publications produce higher extraction accuracy due to better formatting styles and advancement in converters from Word/LaTeX to PDF.

\subsection{Processing time}
\label{sec:processing_time}
\begin{figure}[!thb]
\vspace{-0.4cm}
  \centering
  \resizebox{0.5\linewidth}{!}{\includegraphics{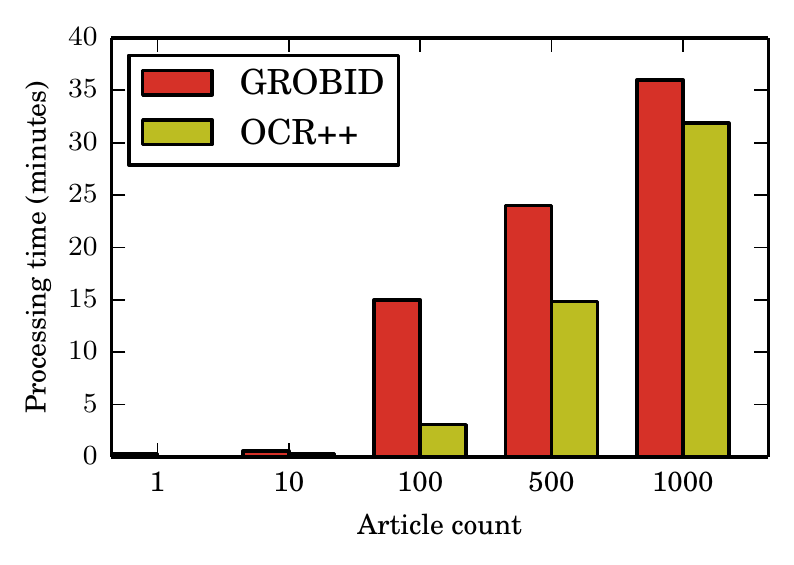}}
  \caption{Comparison between batch processing time of \textit{GROBID} and \textit{OCR++}.} \label{fig:processing_time}
  \vspace{-0.5cm}
\end{figure}

To compare the processing times, we conducted experiments on a set of 1000 PDFs. The evaluation was performed on a single 64-bit machine, eight core, 2003.0 MHz processor and CentOS 6.5 version. Figure~\ref{fig:processing_time} demonstrates comparison between processing time of \textit{GROBID} and \textit{OCR++}, 
while processing some PDF articles in batch mode. There is significant difference in the execution time of \textit{GROBID} and \textit{OCR++},
with \textit{OCR++} being much faster than \textit{GROBID} for processing a batch of 100 articles. 

\subsection{User experience study}
\label{sec:usr_exp}
To conduct a user experience study, we present \textit{OCR++} to a group of researchers (subjects). Each subject is given two URLs: 
1) \text{OCR++} server URL and 2) Google survey 
form\footnote{http://tinyurl.com/juxq2bt}. 
A subject can upload any research article in PDF format on the server and visualize the output. 
In the end, the subject has to fill a response sheet on the Google form. We ask subjects questions related to their experience such as, a) which extraction task did you like the most? b) have you found the system to be really useful? c) have you used similar kind of system before, d) do you find the system slow, fast or moderate, e) comments on the overall system experience and f) drawbacks of the system and suggestions for improvements.

Total 30 subjects participated in the user experience survey. Among the most liked sub-tasks, title extraction comes first with 50\% of votes. Affiliation and author name extraction tasks come second and third respectively. All the subjects found the system to be very useful. Only 
two of the subjects had used a similar system before. As far as the computational speed is concerned, 50\% subjects found the system performance to be fast while 33\% felt it to be moderate. 
\vspace{-0.3cm}
\section{Use Cases}
\label{sec:usecase}
\vspace{-0.1cm}
\subsection{Curation of dataset links}
\vspace{-0.2cm}
With the community is seeing a push towards reproducibility of results, the links to datasets in the research papers are becoming very informative sources for researchers. Nevertheless, to best of our knowledge, we do not find any work on automatic curation of dataset links from the conference proceedings.
With \textit{OCR++}, we can automatically curate dataset related links present in the articles. In order to investigate this in further detail, we aimed to extract dataset links 
from the NLP venue proceedings. We ran \textit{OCR++} on four NLP proceedings, ACL 2015, NAACL 2015, ACL 2014, and  ECAL 2014, available in PDF format. We extract all the URLs present in the proceedings. We then filter those URLs which are either part of \textit{Datasets} 
section's body or are present in the footnotes of \textit{Datasets} section, along with the URLs that consist of one of the three tokens: \textit{datasets, data, dumps}.
Table~\ref{tab:use-case_I_I} presents statistics over these four proceedings for the extraction task. From the dataset links thus obtained, precision was found by human judgement as to whether a retrieved link corresponds to a dataset. One clear trend we saw was the increase in the number of dataset links from year 2014 to 2015. In some cases, retrieved link corresponds to project pages, tools, researcher's homepage etc. resulting in lowering of precision values.

\begin{table}[!thb]
\vspace{-0.3cm}
 \centering
 \caption{Proceedings dataset extraction statistics: 
 Article count represents total number of articles present in the proceedings. 
 Total links and Dataset links correspond to total number of unique URLs and total number of 
 unique dataset links extracted by OCR++ respectively. Precision measures correct number of dataset links.}
 
 \resizebox{0.8\textwidth}{!}{\begin{minipage}{\textwidth}
  \begin{tabular}{cccccc}
  \hline
  Venue &Year&Articles Count&Total links& Dataset links & Precision\\\hline
  ACL&2015&174&345&38&0.74\\
  NAACL&2015&186&186&18&0.50\\
  ACL&2014&139&202&16&0.50\\
  EACL&2014&78&141&12&0.67\\\hline
 \end{tabular}
 \end{minipage}}
\label{tab:use-case_I_I}
\vspace{-0.5cm}
\end{table}
\subsection{Section-wise citation distribution}
\begin{figure}[!thb]
\vspace{-0.2cm}
  \centering
  \resizebox{0.5\linewidth}{!}{\includegraphics{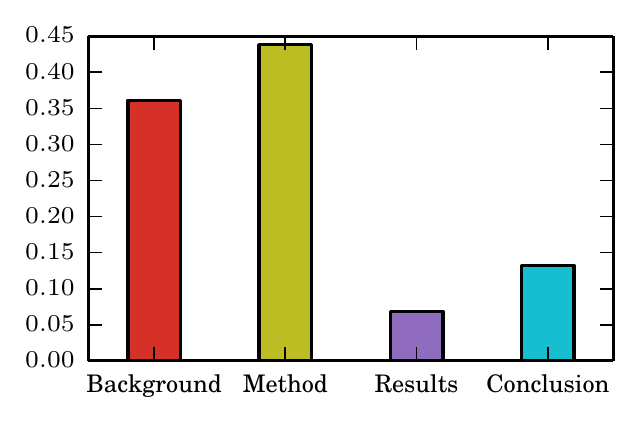}}
  \caption{Sectionwise citation distribution in article dataset} \label{fig:Citation_distribution}
  \vspace{-0.5cm}
\end{figure}

\vspace{-0.2cm}
Citation instance count plays a very important role in determining future popularity of a research paper. An article's text is distributed among several sections. Some sections have more fraction of citations than the rest. In the second use case, we plan to study the section-wise citation distribution. Section-wise citation distribution refers to how citations are distributed over multiple sections in the article's text. This is an important characteristic of the citations and has recently been used for developing a faceted recommendation system \cite{chakraborty2016ferosa}. We group specific sections to 5 generic sections, Background, Datasets, Method, Result/Evaluation and Discussion/Conclusion. 
Table \ref{tab:use-case_II_I} shows an example mapping from specific to generic section names. Note that this mapping can be changed as per the requirement. Figure \ref{fig:Citation_distribution} shows citation distribution for article dataset consisting of the 138 articles mentioned earlier. 
Maximum number of citations are present in the method section, followed by background and discussion and conclusion. Result section comprises the least number of citations. 

\begin{table}[!thb]
\vspace{-0.4cm}
 \centering
\caption{Specific to generic section mapping}
  \resizebox{0.8\textwidth}{!}{\begin{minipage}{\textwidth}
  \begin{tabular}{l|l}
  \hline
  Generic section&Specific sections\\\hline
  Background&Introduction, Related Work, Background\\
  Method& Methodology, \textit{Method Specific names}\\
  Result/Evaluation&Results, Evaluation, Metrics\\
  Discussion/Conclusion& Discussion, Conclusion, Acknowledgment \\\hline
 \end{tabular}
 \end{minipage}}
\label{tab:use-case_II_I}
\vspace{-0.2cm}
\end{table}

\vspace{-0.2cm}
\section{Deployment}
\label{deployment}
\vspace{-0.2cm}
The current version of \textit{OCR++} is deployed at our research group server~\footnote{CNeRG. http://www.cnergres.iitkgp.ac.in}. The present  infrastructure consists of single CentOS instance.~\footnote{OCR++ server. http://www.cnergres.iitkgp.ac.in/OCR++/home/}. We also make entire source code publicly available~\footnote{Source code. http://tinyurl.com/hs9oap2.}.
\vspace{-0.2cm}
\section{Related work}
\label{sec:related}
\vspace{-0.2cm}
Researchers follow several different approaches for individual extraction tasks.
The approaches based on image processing segments document image into several text blocks. Further,
each segmented block is classified into predefined set of logical blocks using machine learning algorithms. 
Gobbledoc~\cite{Nagy:1992:PDI:146970.146971} used X-Y tree data structure that converts the two-dimensional page segmentation
problem into a series of one-dimensional string-parsing problem. \newcite{dengel1995clustering} 
employed the concept language of the GTree for logical labeling. Similar work by \newcite{esposito1995knowledge} presented a
hybrid approach to segment an image by means of a top-down technique and then bottom-up approach to form complex layout component.

Similarly, current state-of-the-art systems use support vector machine (SVM)~\cite{han2003automatic} and conditional random field 
(CRF)~\cite{councill2008parscit,luong2012logical,Lafferty:2001:CRF:645530.655813} based machine learning models for information extraction. 
A study by \newcite{granitzer2012comparison} compares  \textit{Parsecit} (a CRF based system) and 
\textit{Mendeley Desktop}\footnote{https://www.mendeley.com/} (a SVM based system). They observed that SVMs provide 
reasonable performance in solving the challenge of metadata extraction than CRF based approach. However, \newcite{lipinski2013evaluation} observed that \textit{GROBID} (a CRF based system) performed better than \textit{Mendeley Desktop}. 

\vspace{-0.2cm}
\section{Conclusions}
\vspace{-0.2cm}
The goal of this work was to develop an open-source information extraction framework for scientific articles using generic patterns present in various publication formats. In particular, we extract metadata information and section related information. 
The framework also performs two mapping tasks, author and e-mail mapping and citations to reference mapping. Despite OCR errors and 
the great difference in the publishing formats, the framework outperforms the state-of-the-art systems with high margin. 
We find that the hand-written rules and heuristics produced better results than previously proposed machine learning models. 

The current framework has certain limitations. As described in Section \ref{tasks}, we employ \textit{pdf2xml} to convert PDF article into a rich XML file. Even though the XML file consists of rich metadata, it suffers from common errors generated during OCR conversion. Example of such common errors are end-of-line hyphenation and character encoding problem. This is a common problem especially in two-column articles. Secondly, the current version of \textit{pdf2xml} lacks character encoding for non English characters. The two mentioned
major problems along with other minor OCR conversion errors are directly reflected in the \textit{OCR++} output. 

As discussed in previous section, considerable amount of work has already been done to extract reference entities (title, publisher name, date, DOI, etc.) with high accuracy. Therefore, \textit{OCR++} does not aim to extract reference entities. In future, we aim to extend current framework by extracting information present in figures and tables. Figures and tables present concise statistics about dataset and results. We are also currently in process to extend the functionality for non English articles.

\bibliography{acl2016}
\bibliographystyle{acl2016}
\end{document}